\documentclass[twocolumn,preprintnumbers,amsmath,amssymb,prb]{revtex4-1}

\usepackage{graphicx}
\usepackage{dcolumn}
\usepackage{subfigure}
\usepackage{bm}

\begin{document}


\title{On the accuracy of commonly used density functional approximations in determining the elastic constants of insulators and semiconductors}

\author{M. R{\aa}sander}\email{m.rasander@imperial.ac.uk}
\affiliation{%
Department of Materials, Imperial College London, SW7 2AZ, London, United Kingdom
}%
\author{M. A. Moram}
\affiliation{%
Department of Materials, Imperial College London, SW7 2AZ, London, United Kingdom
}%
\date{\today}

\begin{abstract}
We have performed density functional calculations using a range of local and semi-local as well as hybrid density functional approximations of the structure and elastic constants of 18 semiconductors and insulators. We find that most of the approximations have a very small error in the lattice constants, of the order of 1\%, while the error in the elastic constants and bulk modulus are much larger, at about 10\%. In addition, we find that the error in the elastic constants, $c_{ij}$, are larger compared to the error in the bulk modulus. Depending on the functional and which error estimate that is being used, the difference in the error between the elastic constants and the bulk modulus can be rather large, about a factor of two. According to our study, the overall best performing density functional approximation for determining the structure and elastic properties is the PBEsol, closely followed by the two hybrid functionals PBE0 and HSE, and the AM05 functional. 
\end{abstract}

\maketitle
\section{Introduction}

\par
Fundamental material parameters are needed to design, characterize and simulate new devices effectively. For example, accurate elastic constants are essential for determining the composition of epitaxial films using X-ray diffraction,\cite{j} for assessing the critical thicknesses for strain relaxation in device heterostructures\cite{k} and for modeling the behavior of dislocations in these materials.\cite{l} Elastic constants can also be used to obtain the geometry of the individual layers in- and out-of-plane in multilayer structures.\cite{Rasander} However, obtaining elastic constants experimentally is sometimes difficult. For the III-nitrides, for example, it is very difficult to determine the elastic constants accurately through experiment due to the inherently large measurement uncertainties and limitations in the sample quality. Therefore, theoretical elastic constants values are used routinely for these systems.\cite{m} Furthermore, for other semiconductor systems such as many II-IV nitride systems\cite{Paudel-2009,Rasander-Moram} theoretical elastic constants are absolutely necessary since experimental values are lacking. It is therefore necessary to have theoretical tools that are accurate and efficient with the additional requirement to have a predictive power. Density functional theory (DFT) is such a tool and it is now established that the crystal structure, lattice dynamics and electronic structure of almost any element or compound can be accurately treated by this theory. Neglecting the technical and numerical aspects of the implementation, the accuracy in a DF calculation is determined by the exchange-correlation (XC) energy functional. The first approximation for the XC energy functional that was developed is the local density approximation (LDA) which assumes that the XC energy of the electron density in a material is identical to the XC energy of the free electron gas with the same electron density. Even though the LDA is a very simple approximation, it has been found to give remarkably good results for many material properties. However, this level of accuracy is not sufficient for many of the applications that have been mentioned above. Many approximations have however, been proposed in order to improve on the LDA, starting with the generalised gradient approximations (eg. PBE,\cite{PBE} PW91,\cite{PW91,PW91b} revPBE,\cite{revPBE} RPBE,\cite{RPBE} PBEsol\cite{PBEsol}) or other semi-local functionals (eg. AM05\cite{Armiento,Mattsson}), to the more elaborate meta-GGAs (eg. TPSS\cite{TPSS1,Staroverov2004}) and hybrid density functionals (eg. PBE0\cite{PBE0} and HSE\cite{Heyd2003,Heyd2006}). 
\par
The aim with the present study is to establish the level of accuracy of many of the previously mentioned XC energy functional approximations, namely the LDA, PBE, AM05, PBEsol, RPBE, TPSS, PBE0 and HSE approximations, in determining the structure and especially the elastic properties of semiconductors and insulators. Previous comparative studies for solids have mostly focused on the description of the lattice constants\cite{Staroverov2004,Paier2006,Paier2006errata,Krukau2006,Mattsson,Csonka2009,Haas2009b,Haas2009a} and the bulk modulus \cite{Staroverov2004,Paier2006,Paier2006errata,Mattsson,Csonka2009} for both metallic and semiconducting/insulating systems. We will discuss both the accuracy in the lattice constants and the bulk modulus. Furthermore, we will discuss the accuracy in determining the single crystal elastic constants, $c_{ij}$. The bulk modulus is an important physical property of a material. It measures the response of a material under hydrostatic compression. However, it contains less information than the knowledge of the elastic constants. It is therefore important to assess the accuracy of DF calculations in determining the elastic constants in relation to the accuracy in determining the structural properties and the bulk modulus. Especially, for the applications listed previously where the individual elastic constants are absolutely essential.
\par
The 18 systems that have been investigated in this work include elemental semiconductors (C, Si and Ge), zinc-blende semiconductors (BN, BP, GaP, GaAs, InP, InAs, InSb, and SiC) and insulators in the rock-salt structure (LiF, LiCl, NaF and MgO). Additionally, we have included two systems in the fluorite structure (CaF$_{2}$ and Mg$_{2}$Si) as well as the small band gap skutterudite CoSb$_{3}$. These latter systems have been included to provide more structural complexity compared to the traditional zinc-blende and rock-salt systems. Especially, CoSb3 has a large unit cell containing 16 atoms, where large voids are present in the structure, see for example Ref.~\onlinecite{Rasander-FeSb3} and reference therein.
\par
The paper is outlined as follows: In Section \ref{sec:theory} we will give a brief theoretical background on the DF approximations which have been used in this study. For more details, we refer to the original publications. In Section \ref{sec:details} we present the details of our calculations and in Section \ref{sec:results} we will present our results. Finally, in Section \ref{sec:conclusions} we will summarise our findings and present the conclusions.

\begin{table*}[t]
\caption{\label{tab:error} Mean error (ME), mean absolute error (MAE), root mean square error (RMSE) and mean absolute relative error (MARE) for lattice constants, $a$,  elastic constants, $c_{ij}$, averaged over all combinations of $ij$ (i.e. $ij=11, 12$ and 44), and for the bulk modulus, $B$, for all functionals used in this work. The values shown in bold are the measures that show the closest agreement with the experimental values.}
\begin{ruledtabular}
\begin{tabular}{lccccccccccccccc}
 & \multicolumn{5}{c}{$a$ (\AA)} & \multicolumn{5}{c}{$c_{ij}$ (GPa)}  & \multicolumn{5}{c}{$B$ (GPa)} \\
\cline{2-6}\cline{7-11}\cline{12-16}
XC & ME & MAE & RMSE & MRE & MARE  & ME & MAE & RMSE & MRE & MARE &ME & MAE & RMSE & MRE & MARE  \\
\hline
LDA & -0.047 & 0.047 & 0.061 & -0.9\% & 0.9\% & 7.1 & 9.2 & 15.9 & 5.9\% & 8.1\% & 7.2 & 7.9 & 10.9 & 8.0\% & 8.8\%\\
PBE & 0.076 & 0.076 & 0.086 & 1.4\% & 1.4\% & -8.7 & 11.6 & 14.7 & -8.7\% & 10.8\% & -8.9 & 9.0 & 11.1 & -8.9\% & 8.9\%\\
AM05 & 0.024 & 0.033 & 0.044 & 0.5\% & 0.6\% & -2.9 & 7.8 & \textbf{11.5} & -4.3\% &  7.4\% & -3.1 & 5.9 & 7.0 & -4.1\% &  6.3\%\\
PBEsol & \textbf{0.013} & \textbf{0.025} & \textbf{0.031} & \textbf{0.3\%} &  \textbf{0.5\%} & \textbf{-1.5} & \textbf{7.4} & 11.6 & \textbf{-2.1\%} & 6.8\% & \textbf{-0.9} & 5.1 & \textbf{6.5} & \textbf{-0.7\%} & 5.3\%\\
RPBE & 0.135 & 0.135 & 0.149 & 2.6\% & 2.6\% & -15.6 & 17.2 & 21.8 & -15.2\% & 16.1\% & -16.2 & 16.2 & 18.1 & -16.6\% & 16.6\%\\
TPSS & 0.052 & 0.052 & 0.061 & 1.0\% & 1.0\% & -6.1 & 10.1 & 13.5 & -6.2\% & 9.1\% & -6.6 & 7.7 & 9.2 & -5.9\% & 7.8\%\\
PBE0 & 0.023 & \textbf{0.025} & 0.033 & 0.4\% & \textbf{0.5\%} & 6.6 & 8.4 & 16.3 & 3.0\% & 5.5\% & 4.0 & 4.5 & 8.2 & 2.4\% & 3.2\%\\
HSE & 0.025 & 0.030 & 0.039 & 0.5\% & 0.6\% & 5.9 & 8.2 & 16.0 & \textbf{2.1\%} & \textbf{5.3\%} & 3.1 & \textbf{4.1} & 8.0 & 1.2\% & \textbf{2.7\%}\\
\end{tabular}
\end{ruledtabular}
\end{table*}

\section{Theoretical background}\label{sec:theory}
Within the Kohn-Sham approach to density functional theory, the total energy of a system of electrons is given by (all equations are given in atomic units)
\begin{multline}
E_{tot} = T_{s} + \int v_{ext}({\bm r})n({\bm r})d^3r\\
  +\frac{1}{2}\int\int\frac{n({\bm r})n({\bm r}')}{|{\bm r}-{\bm r}'|}d^3r d^3r' + E_{xc},
\end{multline}
where $T_{s}$ is the kinetic energy of a system of noninteracting electrons and the following terms represent the electron-nucleus energy, electron-electron energy and $E_{xc}$ is the exchange-correlation (XC) energy which is unknown and has to be approximated as will be discussed below. This term may be divided further into an exchange and a correlation term: $E_{xc}=E_{x}+E_{c}$. 
\par
\subsection{Local and semi-local functionals}
For a local or semi-local density functionals, the XC energy functional can, in general, be described by 
\begin{multline}\label{eq:XC-energy}
E_{xc}\bigl[n_{\uparrow}, n_{\downarrow} \bigr] \\
=\int d^3r\,n({\bm r})\epsilon_{xc}\bigl(n_{\uparrow}, n_{\downarrow}, \nabla n_{\uparrow}, \nabla n_{\downarrow}, \tau_{\uparrow},\tau_{\downarrow} \bigr),
\end{multline}
where $n_{\alpha}$, $\nabla n_{\alpha}$ and $\tau_{\alpha}$ are the electron density, gradient of the density and the kinetic energy density of spin $\alpha=\uparrow$~or~$\downarrow$, respectively. However, we will from now on neglect the spin index. Within the local density approximation (LDA) the XC energy only depends on the electron density and can be expressed as
\begin{equation}
E_{xc}^{\text{LDA}}[n] = \int n({\bm r})\epsilon_{xc}^{\text{LDA}}\bigl(n({\bm r})\bigr)d^3r,
\end{equation}
where the XC energy per unit volume $\epsilon_{xc}^{\text{LDA}}$ is a function of the electron density $n({\bm r})$ and is chosen to be identical to the XC energy of the uniform electron gas with the same density. The exchange part is given by $\epsilon_{xc}^{\text{LDA}}=-(3/4)(3/\pi)^{1/3}n^{4/3}$. The correlation energy is obtained through a fit\cite{PerdewLDA} of accurate quantum Monte Carlo calculations for the uniform electron gas.\cite{Ceperley}  The LDA gives reasonably reliable geometries for solids and for elastic constants the typical error has in previous studies been found to be in the order of 5-10\%\cite{Fast} depending on the system. However, it fails badly for the atomisation energies of molecules and solids. 
\par
The generalized gradient approximation (GGA) constitute a family of approximations where the XC energy can be expressed as 
\begin{multline}
E_{xc}^{\text{GGA}}[n] = \int n({\bm r})\epsilon_{xc}^{\text{GGA}}\bigl(n({\bm r}), \nabla n({\bm r})\bigr)d^3r,\\
= \int n({\bm r})\epsilon_{xc}^{\text{LDA}}\bigl(r_{s}({\bm r})\bigr)F_{xc}\bigl(r_s({\bm r}), s({\bm r})\bigr)d^3r,
\end{multline}
where $F_{xc}(r_{s},s)=F_{x}(s)+F_{c}(r_{s},s)$ is the enhancement factor, $r_{s}^3 = 3/(4\pi n)$ is the Wigner-Seitz radius and $s=|\nabla n|/[2(3\pi^2)^{1/3}n^{4/3}]$ is the reduced density gradient. In the literature two types of GGA functionals can be found: (i) the empirical functionals, whose parameters were determined by fitting to experimental or first principles data, and (ii) the parameter free functionals, e.g. PBE\cite{PBE} and PW91\cite{PW91,PW91b}, whose parameters were determined in order to satisfy mathematical relations which are known to hold for the exact functional. Within this context it should be mentioned that the parameter free functionals contain arbitrary choices such as the analytical form chosen to represent $F_{xc}$ or the choice of constraints to be satisfied. Here we will only discuss parameter free functionals.
\par
The PBE functional\cite{PBE} was designed to satisfy several conditions obeyed by the exact functional, e.g. the correct uniform electron gas limit (i.e. LDA is recovered when $s=0$), the Lieb-Oxford bound\cite{Lieb-Oxford} ($E_{x}\geq E_{xc}\geq-1.679\int n^{4/3}d^3r$) and the LDA linear response. The enhancement factor for exchange is given by
\begin{equation}
F_{x}^{\text{PBE}}(s) = 1+\kappa-\frac{\kappa}{1+\frac{\mu}{\kappa}s^2},
\end{equation}
where $\kappa=0.804$ and $\mu=0.21951$. One advantage of the PBE functional is that it performs well for finite and infinite systems and therefore rather diverse in its applications. For solids, the PBE has a tendency of underbinding, i.e. yielding too large binding distances and too small formation energies, and several functionals have been designed to improve on the performance of the PBE for solids. Here we will briefly mention two, namely PBEsol\cite{PBEsol} and RPBE\cite{RPBE}.
\par
The PBEsol\cite{PBEsol} was designed to be more accurate than PBE for solids and surfaces. Within a GGA there is a choice to be made between the accuracy in e.g. lattice parameters and surface energies in comparison to accurate atomisation energies. The PBE, as has been mentioned, performs well for both solids and molecules, however, an improved functionality in one area of applications, e.g. for solids, will worsen the performance for atoms and molecules. Within the PBEsol the same analytic form as in the PBE is used, with some of the parameters changed to satisfy modified constraints in comparison to PBE.\cite{PBEsol} The PBEsol has indeed been found to yield improved lattice parameters and surface energies compared to PBE, however, the performance for atoms and molecules are worse than for the PBE. 
\par
In order to achieve a XC functional with improved adsorption energies compared to LDA and PBE, Hammer {\it et al.}\cite{RPBE} proposed a slightly adapted form of the exchange enhancement factor, $F_{x}$, compared to the PBE, with
\begin{equation}
F_{x}^{\text{RPBE}}(s) = 1+\kappa\bigl( 1-e^{-\mu s^2/\kappa} \bigr),
\end{equation}
with an identical value for $\kappa$ and all correlation terms identical as in PBE. It was found that adsorption energies of several atoms and molecules on metallic surfaces was greatly improved with the RPBE.\cite{RPBE} However, lattice constants are overestimated in RPBE and the bulk modulus is underestimated significantly, which is related to the strong $s$ dependence compared to the PBE.\cite{Mattsson}
\par
While the standard procedures for developing new GGA functionals are to fit parameters to experimental data and/or by satisfying universal mathematical conditions on the exact DF, the AM05 functional was the first functional to use a subsystem functional scheme.\cite{Armiento} The idea is to use separate functionals from different model systems for which the XC energy is known, namely the uniform electron gas and the surface jellium. A DF index, which depends on the reduced density gradient, is then used to locally determine the nature of the system.\cite{Armiento,Mattsson} AM05 is therefore a systematic improvement over LDA, by the inclusion of terms that depend on the density gradient while maintaining the XC limit of the LDA. It has been found that the PBEsol and AM05 have very similar performance for many systems, with the PBEsol performing slightly better than AM05.\cite{Csonka2009} This was argued by Csonka {\it et al}\cite{Csonka2009} to be due to the very similar behaviour of the XC enhancement factor $F_{xc}(s,r_{s})$ between PBEsol and AM05 in solids where $s<1$ everywhere. For some solids with $s_{max}>>1$ the difference between PBEsol and AM05 is greater.\cite{Csonka2009}
\par
If the dependence of the electron density, gradient of the density and the kinetic energy density are all used in the construction of the XC energy functional as described in Eq. (\ref{eq:XC-energy}) we have a so called meta-GGA XC energy functional. Compared to a GGA functional the inclusion of the kinetic energy adds extra flexibility in the construction of a XC energy functional. Such functionals has the potential to correctly treat effects that cannot be treated accurately in LDA and GGA. Several types of meta-GGA functionals have been presented. Here, we will use the TPSS meta-GGA functional,\cite{TPSS1} which satisfies several exact constrains without using any empirical parameters and it has been found to be reliable for both molecules and solids.\cite{TPSS1,Staroverov2004}

\subsection{Hybrid density functionals}

Hybrid DF are non-local theories where some of the exchange part of a standard XC functional, e.g. PBE, has been substituted by some amount of Hartree-Fock exchange energy. The two hybrid functionals that have been used in this study, PBE0\cite{PBE0} and HSE,\cite{Heyd2003,Heyd2006} both have 25\% Hartee-Fock exchange energy substituting regular PBE exchange. In the PBE0 approximation the XC energy is expressed by
\begin{equation}
E_{xc}^{\rm PBE0} = \frac{1}{4}E_{x}^{\rm HF} + \frac{3}{4}E_{x}^{\rm PBE} + E_{c}^{\rm PBE},
\end{equation}
where $E_{x}^{\rm PBE}$ and $E_{c}^{\rm PBE}$ are the exchange and correlation terms in the PBE approximation and $E_{x}^{\rm HF}$ is the Hartree-Fock exchange energy. In the HSE approximation the XC energy is given by\cite{Heyd2003,Heyd2006}
\begin{equation}\label{eq:HSE1}
E_{xc}^{\rm HSE} = E_{x}^{\rm HSE} + E_{c}^{\rm HSE},
\end{equation}
where
\begin{multline}
E_{x}^{\rm HSE} = \alpha E_{x}^{\rm HF,SR}(\omega) + (1-\alpha)E_{x}^{\rm PBE,SR}(\omega)\\
+ E_{x}^{PBE,LR}(\omega),
\end{multline}
where LR and SR denote the long range and short range parts of the exchange energy respectively, see Refs.~\onlinecite{Heyd2003} and \onlinecite{Heyd2006}. $\alpha$ is a mixing parameter governing the amount of the non-local Hartree-Fock exchange and $\omega$ is a screening parameter that controls the spatial range over which the non-local exchange part is important. The correlation energy in the HSE approximation, $E_{c}^{\rm HSE}$, is taken to be identical to the PBE correlation energy as in the PBE0 approximation. The amount of Hartree-Fock exchange in the HSE is identical to the amount in PBE0, i.e. $\alpha = 1/4$. The range separation parameter, $\omega$, on the other hand has to be determined by comparison with experimental data. It has been found that $\omega =0.2-0.3$~\AA$^{-1}$ gives good results with regards to structural as well as electronic properties of materials, with $\omega = 0.2$~\AA$^{-1}$ being the optimum choice.\cite{Heyd2003,Heyd2006,Matsushita2011} We have in this study used $\omega = 0.2$~\AA$^{-1}$. Note that for $\omega=0$ Eq.~(\ref{eq:HSE1}) is equivalent to the PBE0 and, in addition, Eq.~(\ref{eq:HSE1}) asymptotically reaches the PBE for $\omega\longrightarrow\infty$.\cite{Heyd2003,Rasander-Cu2Se} Hybrid DF are more accurate than LDA and PBE for many properties. Especially, it is possible to obtain reliable band gaps using hybrid theories. However, the non-local nature of these approximations make them computationally very expensive. In addition, it has been found that PBEsol and AM05 are just as accurate as hybrid approximations for structural properties and in determining the bulk modulus.\cite{Mattsson}
\begin{figure}[t]
\includegraphics[width=8.5cm]{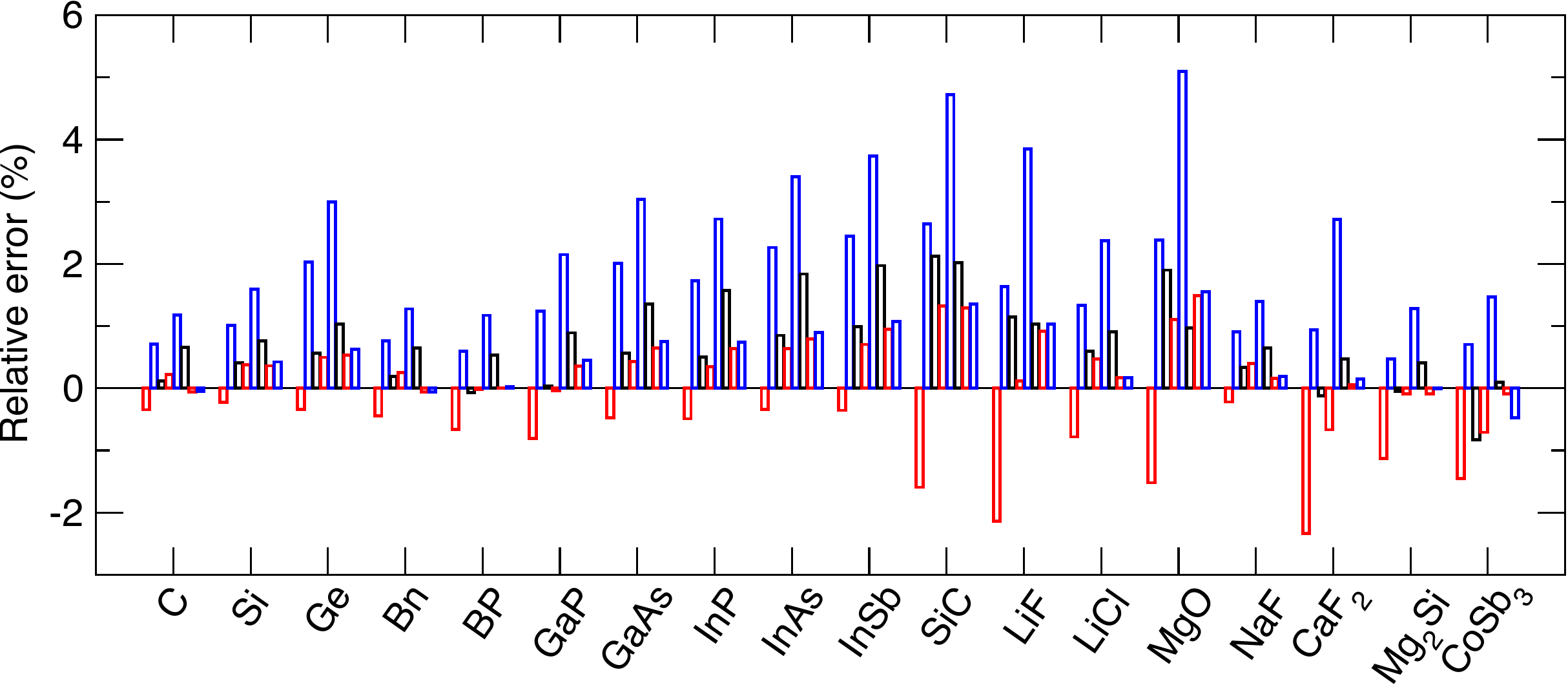}
\caption{\label{fig:lattice-constants} (Color online) Calculated relative error in the lattice constants, $a$, $\text{RE}=(a-a_{0})/a_{0}$, where $a_{0}$ is the experimental reference. The bars are the relative errors for (from left to right) the LDA, PBE, AM05, PBEsol, RPBE, TPSS, PBE0 and HSE approximations.}
\end{figure}
\begin{figure}[t]
\includegraphics[width=8.5cm]{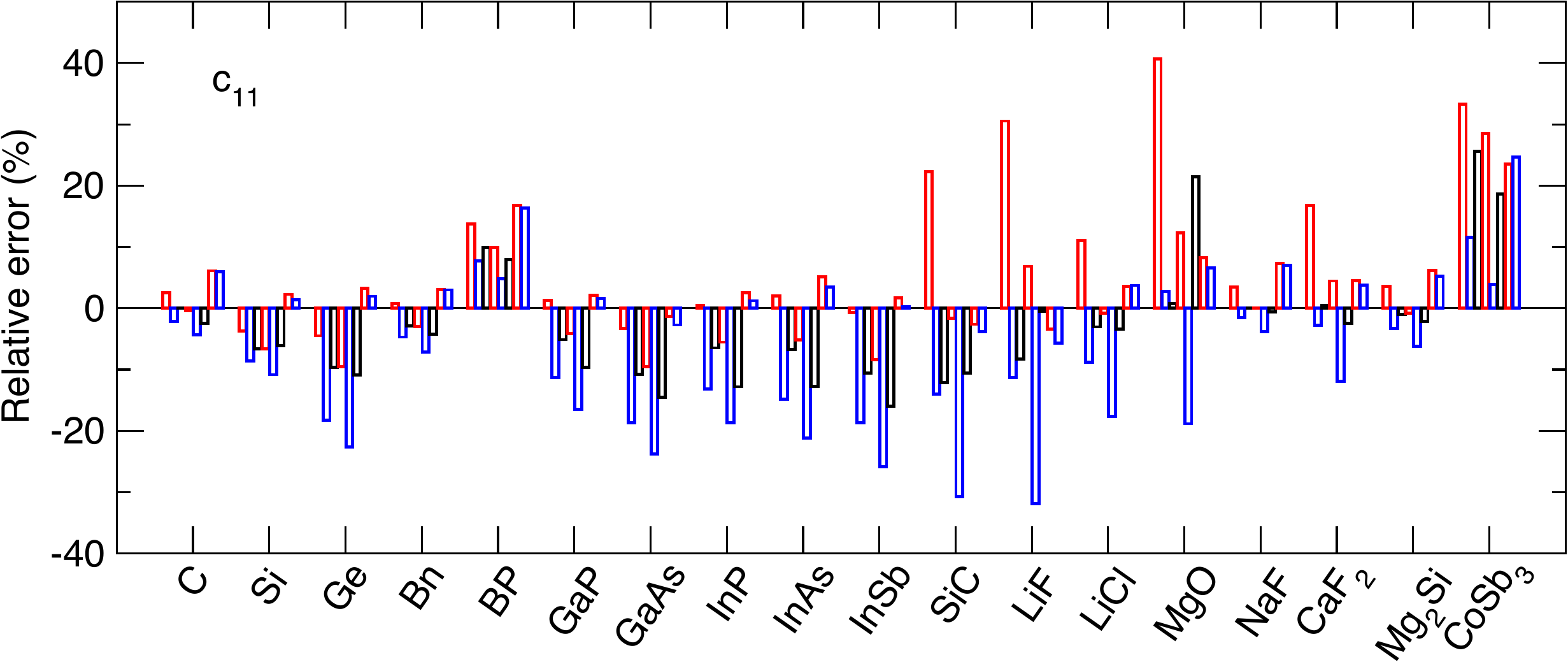}
\caption{\label{fig:c11} (Color online) Calculated relative error in $c_{11}$, $\text{RE}=(c_{11}-c_{11}^{0})/c_{11}^{0}$, where $c_{11}^{0}$ is the experimental reference. The bars are the relative errors for (from left to right) the LDA, PBE, AM05, PBEsol, RPBE, TPSS, PBE0 and HSE approximations.}
\end{figure}
\begin{figure}[t]
\includegraphics[width=8.5cm]{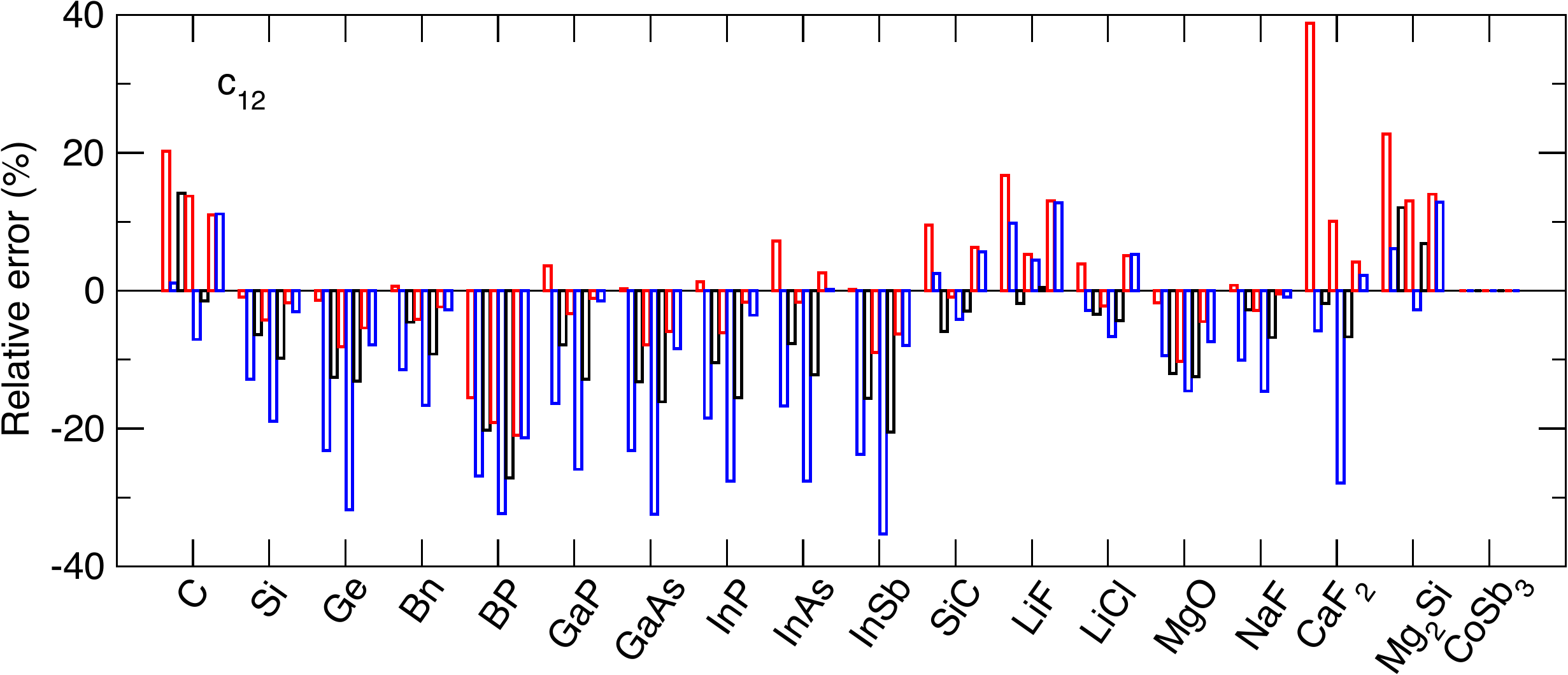}
\caption{\label{fig:c12} (Color online) Calculated relative error in $c_{12}$, $\text{RE}=(c_{12}-c_{12}^{0})/c_{12}^{0}$, where $c_{12}^{0}$ is the experimental reference. The bars are the relative errors for (from left to right) the LDA, PBE, AM05, PBEsol, RPBE, TPSS, PBE0 and HSE approximations. Note that there is no experimental $c_{12}$ data for CoSb$_{3}$.}
\end{figure}
\begin{figure}[t]
\includegraphics[width=8.5cm]{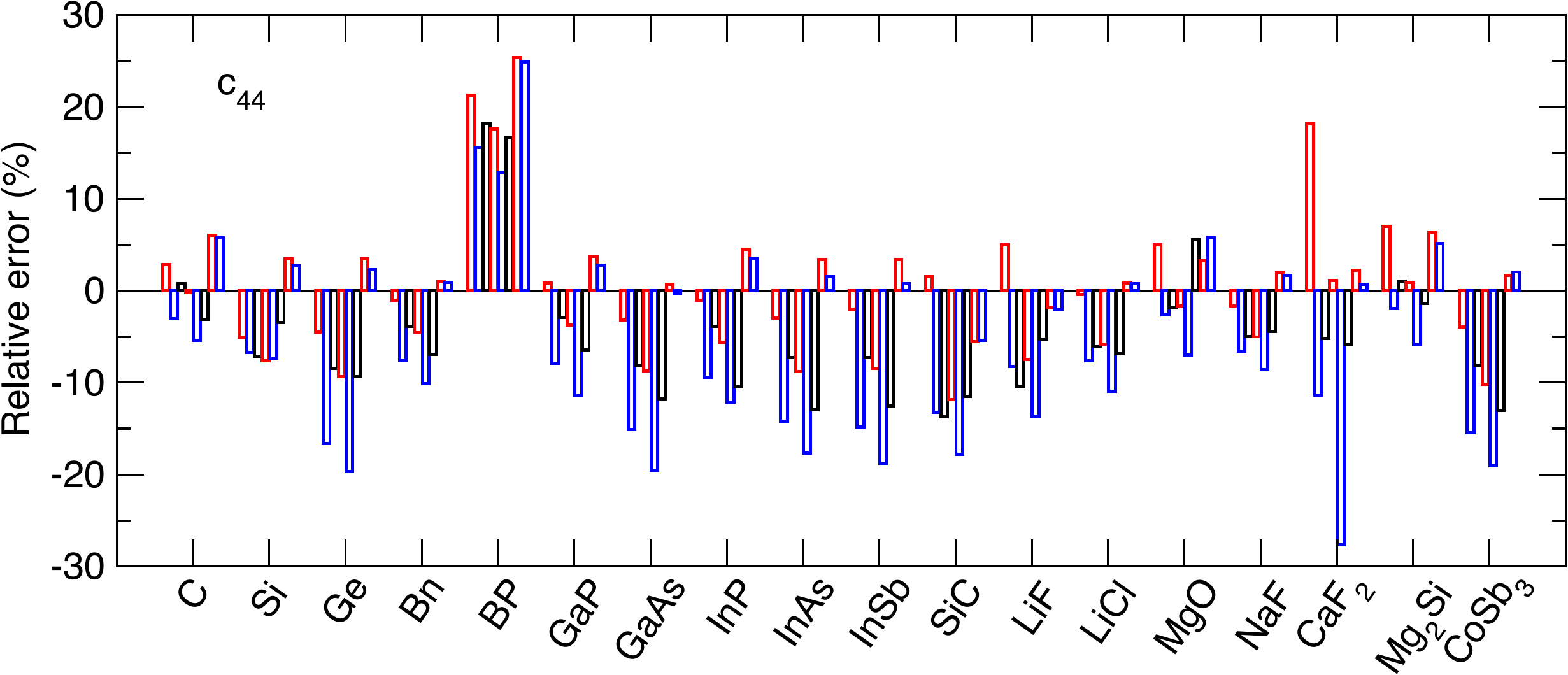}
\caption{\label{fig:c44} (Color online) Calculated relative error in $c_{44}$, $\text{RE}=(c_{44}-c_{44}^{0})/c_{44}^{0}$, where $c_{44}^{0}$ is the experimental reference. The bars are the relative errors for (from left to right) the LDA, PBE, AM05, PBEsol, RPBE, TPSS, PBE0 and HSE approximations.}
\end{figure}
\begin{figure}[t]
\includegraphics[width=8.5cm]{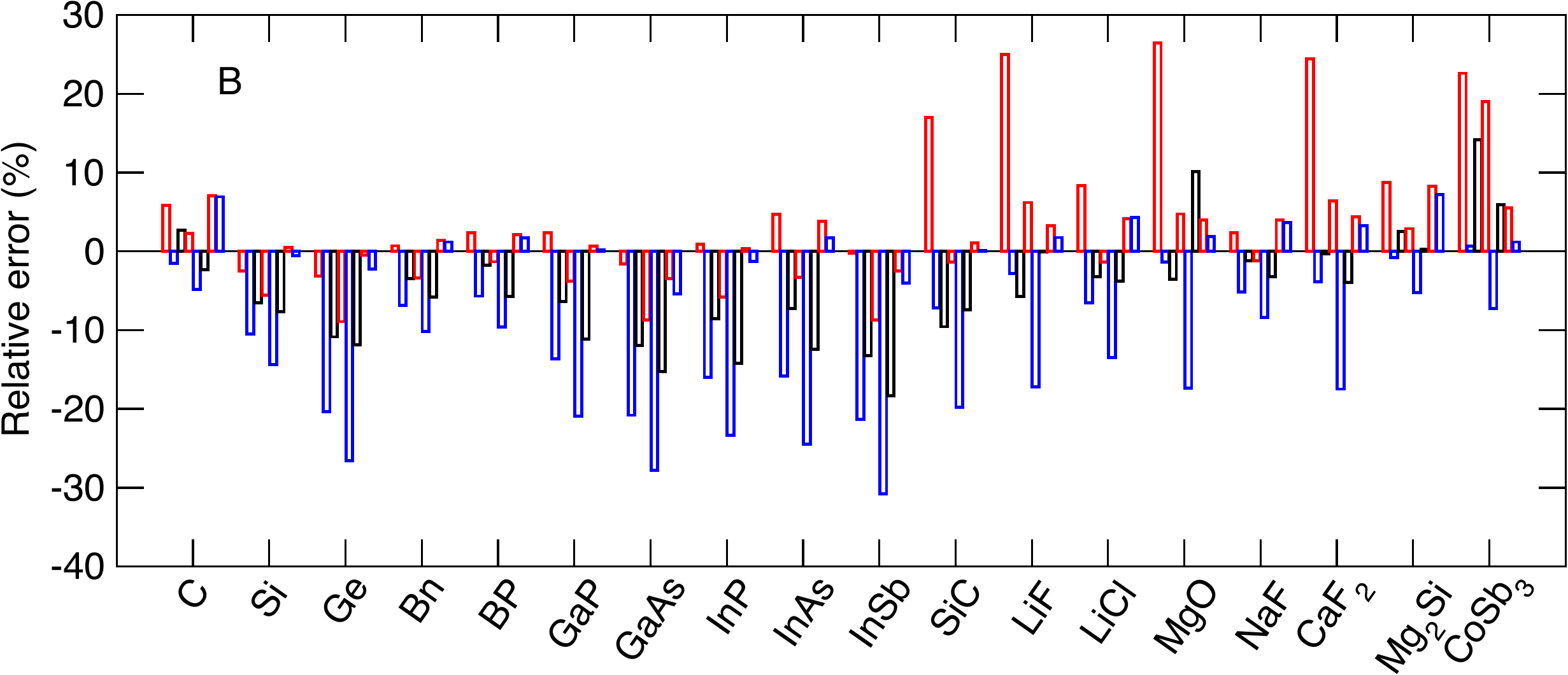}
\caption{\label{fig:B} (Color online) Calculated relative error in the bulk modulus, $B$, $\text{RE}=(B-B_{0})/B_{0}$, where $B_{0}$ is the experimental reference. The bars are the relative errors for (from left to right) the LDA, PBE, AM05, PBEsol, RPBE, TPSS, PBE0 and HSE approximations.}
\end{figure}

\section{Details of the calculations}\label{sec:details}
Density functional calculations have been performed using the projector augmented wave (PAW) method\cite{Blochl} as implemented in the Vienna {\it ab initio} simulation package (VASP).\cite{KresseandFurth,KresseandJoubert} The plane wave energy cut-off was set to 800~eV. For the local and semi-local density functional approximations (LDA, PBE, AM05, RPBE and PBEsol) we have used $\Gamma$-centered k-point meshes with the smallest allowed spacing between k-points of 0.1~\AA$^{-1}$ and for the much more computationally demanding hybrid density functionals (PBE0 and HSE) the spacing was set to 0.4~\AA$^{-1}$. For the meta-GGA TPSS functional we have used 0.1~\AA$^{-1}$. The atomic positions and simulation cell shapes were relaxed until the Hellmann-Feynman forces acting on atoms were smaller than 0.001~eV/\AA.
\par
Apart from when using the LDA, the PAW core potentials used in the calculations were obtained using the PBE functional. For most of the approximations, this is the most obvious choice since the approximations are extensions or revisions of the PBE. The only exception is AM05, but it has been shown\cite{Mattsson} that the difference between using LDA and PBE PAW core potentials in an AM05 calculation is small. For most elements in our study the PAW core potentials were set up using obvious core-valence partitions. The exceptions were Ca (3s$^{2}$3p$^{6}$4s$^{2}$), Ga (3d$^{10}$4s$^{2}$4p$^{1}$), Ge (3d$^{10}$4s$^{2}$4p$^{2}$), In (4d$^{10}$5s$^{2}$5p$^{1}$) and Na (2s$^{2}$2p$^{6}$3s$^{1}$), where semi-core states have been treated as valence states. 
\par
The elastic constants were evaluated following Refs.~\onlinecite{LePage} and \onlinecite{Wu} where the stress is evaluated from the application of a strain to the system and the elastic constants are evaluated from Hooke's law, ${\bm \sigma}=\bar{\bm C}{\bm \epsilon}$, where ${\bm \sigma}$ is the stress tensor, ${\bm \epsilon}$ is the strain tensor and $\bar{\bm C}$ is the elastic constants tensor. This approach in combination with semi-local XC functionals has been used to calculate the elastic constants of both zinc-blende\cite{n} and wurtzite\cite{o,p} group III-nitride alloys, producing accurate elastic constants with a relatively low computational cost.\cite{LePage,o,r} 
\par
In order to assess our calculated data, these have been compared to available experimental data. Lattice and elastic constants have, as much as possible, been extracted from low temperature experiments, ideally extracted from $T=0$~K. However, this is not always possible in which case experimental values taken at room temperature have been used. For the lattice constants, we have whenever possible used the same values as Mattsson {\it et al.}\cite{Mattsson} where the same policy was used. In the case of CoSb$_{3}$ we were unable to find any measured value for the $c_{12}$ elastic constant and our calculated values for $c_{12}$ presented here should be regarded as predictions.
\par
The accuracy of the calculated results have been evaluated by the mean error (ME), the mean absolute error (MAE), the root mean square error (RMSE), the mean relative error (MRE) and the mean absolute relative error (MARE). The ME is given by
\begin{equation}\label{eq:ME}
\text{ME} = \frac{1}{n}\sum_{i=1}^{n}\bigl(\hat{Y}_{i} -  Y_{i}\bigr),
\end{equation}
where $\hat{Y}_{i}$ and $Y_{i}$ are the calculated and experimental values for the quantity of interest, for example the lattice, $a$, or elastic, $c_{ij}$, constants, respectively, and $n$ is the number of systems. Furthermore, the MAE is given by
\begin{equation}\label{eq:MAE}
\text{MAE} = \frac{1}{n}\sum_{i=1}^{n}\bigl|\hat{Y}_{i} -  Y_{i}\bigr|,
\end{equation}
the RMSE is given by
\begin{equation}\label{eq:RMSE}
\text{RMSE} = \sqrt{\frac{1}{n}\sum_{i=1}^{n}\bigl|\hat{Y}_{i} -  Y_{i}\bigr|^2},
\end{equation}
the MRE is given by
\begin{equation}\label{eq:MRE}
\text{MRE} = \frac{1}{n}\sum_{i=1}^{n}\frac{\bigl(\hat{Y}_{i} -  Y_{i}\bigr)}{Y_{i}},
\end{equation}
and the MARE is given by
\begin{equation}\label{eq:MARE}
\text{MARE} = \frac{1}{n}\sum_{i=1}^{n}\frac{\bigl|\hat{Y}_{i} -  Y_{i}\bigr|}{Y_{i}}.
\end{equation}
It will be shown that the choice of the error estimate will affect the conclusions slightly. Note that functionals that consistently over- or underestimate either the lattice constants or the elastic constants will have $|\text{ME}|=|\text{MAE}|$ and $|\text{MRE}|=|\text{MARE}|$.

\section{Results and discussion}\label{sec:results}
In Table~\ref{tab:error} we show the evaluated error measures for the lattice constants, elastic constants and the bulk modulus for the 18 solids in our study. Note that in Table~\ref{tab:error} the error estimates in the three elastic constants are averaged, i.e. $\Delta\bar{c}_{ij}=\bigl(n_{1}\Delta c_{11}+n_{2}\Delta c_{12}+n_{3}\Delta c_{44} \big)/(n_{1}+n_{2}+n_{3})$, where $\Delta c_{ij}$ is any of the measures ME, MAE, MRE or MARE, and $n_{i}$ is the number of evaluated systems for the elastic constant $c_{ij}$. Furthermore, note that the RMSE needs a slightly different treatment, which is apparent from the form of the RMSE in Eq.~(\ref{eq:RMSE}). Experimental references and the calculated results are shown in Tables~\ref{tab:lattice-constants}, \ref{tab:elastic_cubic_c11}, \ref{tab:elastic_cubic_c12}, \ref{tab:elastic_cubic_c44} and \ref{tab:Bmod_cubic}. In addition, Figs. \ref{fig:lattice-constants} through \ref{fig:B} show the relative error in the lattice constants, the elastic constants (one figure for each of the $c_{11}$, $c_{12}$ and $c_{44}$) as well as the bulk modulus for each of the 18 systems in our study.
\par
\begin{table*}[t]
\caption{\label{tab:lattice-constants} Lattice constants, $a$, for systems with cubic symmetry. The experimental lattice constants without the ZPAE subtracted are given within parenthesis. Data shown in bold show the least deviation from experimental values. All units in \AA.}
\begin{ruledtabular}
\begin{tabular}{lccccccccc}
 & Expt.\cite{Landolt-Bornstein} & LDA & PBE & AM05 & PBEsol & RPBE & TPSS & PBE0 & HSE\\
\hline
C & 3.545 (3.567) & 3.533 & 3.570 &  3.549 & 3.553 & 3.587 & 3.568 & 3.543 & 3.543\\
Si  & 5.415 (5.430) &  5.403 & 5.470 & 5.437 & 5.436 & 5.501 & 5.456 & 5.435 & 5.438\\
Ge  & 5.646 (5.658) & 5.627 & 5.761 & 5.678 & 5.674 & 5.815  & 5.705 & 5.676 & 5.682\\
BN & 3.597 (3.615) & 3.581 & 3.624 & 3.604 & 3.606 & 3.587 & 3.620 & 3.595 & 3.595 \\
BP  & 4.522 (4.538) & 4.492 & 4.549 & 4.519 & 4.521 & 4.575  & 4.546 & 4.522 &  4.523\\
GaP  & 5.439 (5.451) & 5.395 & 5.506 & 5.441 & 5.437 & 5.556 & 5.487 & 5.458 & 5.463\\
GaAs & 5.637 (5.648) & 5.610 & 5.751 & 5.669 & 5.661 & 5.808 & 5.714 & 5.674 & 5.680\\
InP  & 5.855 (5.869) & 5.827 & 5.957 & 5.885 & 5.876 & 6.015 & 5.948 & 5.893 & 5.899\\
InAs  & 6.048 (6.058) & 6.027 & 6.185 & 6.099 & 6.086 & 6.254 & 6.159 & 6.096 &  6.103\\
InSb  & 6.473 (6.479) & 6.450 & 6.631 & 6.537 & 6.518 & 6.714 & 6.600 & 6.534 & 6.542\\
SiC  & 4.341 (4.358) & 4.332 & 4.380 & 4.356 & 4.359 & 4.402 & 4.369 & 4.348 &  4.349\\
LiF  & 3.969 (4.010) & 3.906 & 4.074 & 4.054 & 4.022 & 4.157 & 4.049 &  4.021 & 4.023\\
LiCl  & 5.071 (5.106) & 4.963 & 5.154 & 5.129 & 5.077 &  5.266 & 5.123 & 5.118 & 5.124\\
MgO  & 4.186 (4.207) & 4.153 & 4.242 & 4.211 & 4.206 & 4.285  & 4.223 & 4.193 & 4.193\\
NaF  & 4.577 (4.609) & 4.507 & 4.686 & 4.664 & 4.628 & 4.810 &  4.621  & 4.645 & 4.648 \\
CaF$_{2}$  & 5.443 (5.466) & 5.316 & 5.495 & 5.437 & 5.407 & 5.591  & 5.469 &  5.447 & 5.452\\
Mg$_{2}$Si  & 6.329 (6.347) & 6.258 & 6.359 & 6.326 & 6.324 & 6.411 & 6.355 &  6.324 & 6.329\\
CoSb$_{3}$\cite{CoSb3-lattice}  & 9.041 (9.055) & 8.910 & 9.105 & 8.966 & 8.977 & 9.174 & 9.050 & 9.033 & 8.998\\
\\
ME & -& -0.045 & 0.077 & 0.025 & \textbf{0.014} & 0.136 & 0.053 & 0.025 & 0.026 \\
MAE & -& 0.047 & 0.077 & 0.036 & \textbf{0.027} & 0.136 & 0.053 & 0.028 & 0.033 \\
RMSE & -& 0.061 & 0.089 & 0.047 & \textbf{0.034} & 0.152 & 0.064 & 0.037 & 0.042 \\
MRE & - & -0.9\% & 1.4\% & 0.5\% & \textbf{0.3\%} & 2.6\% & 1.0\% & 0.4\% & 0.5\% \\
MARE & -& 0.9\% & 1.5\% & 0.7\% & \textbf{0.5\%} & 2.6\% & 1.0\% & \textbf{0.5\%} & 0.6\%\\
 \end{tabular}
\end{ruledtabular}
\end{table*}

\subsection{Lattice constants}
Experimental lattice constants are usually measured at room temperature and, furthermore, they also include zero-point phonon effects (ZPPE). Therefore, these experimental values are not directly comparable with the results of ground-state density functional calculations since the theory are obtained at $T=0$~K. For lattice constants the ZPPE manifest as a zero-point anharmonic expansion (ZPAE) of the lattice.\cite{Alchagirov2001} This effect can be estimated by\cite{Alchagirov2001}
\begin{equation}\label{eq:correctionlattice}
\frac{\Delta a_{0}^{\text{expt}}}{a_{0}^{\text{expt}}} = \frac{\Delta V_{0}^{\text{expt}}}{3V_{0}^{\text{expt}}} = \frac{3}{16}\bigl(B_{1}-1\bigr)\frac{k_{B}\Theta_{D}}{B_{0}v_{0}^{\text{expt}}},
\end{equation} 
and the ZPAE corrected experimental value is $a_{0}^{\text{expt}}-\Delta a_{0}^{\text{expt}}$. In Eq.~(\ref{eq:correctionlattice}), $v_{0}^{\text{expt}}$ is the experimental volume per atom, $B_{1}$ is the the pressure derivative of the bulk modulus at equilibrium, $B_{0}$ is the bulk modulus at equilibrium, and $\Theta_{D}$ is the Debye temperature. For this correction we have, as done previously in the literature,\cite{Haas2009b,Staroverov2004} used experimental values for $B_{0}$ and $\Theta_{D}$, while we have used theoretical values for $B_{1}$. In Ref.~\onlinecite{Staroverov2004} and \onlinecite{Haas2009b} theoretical $B_{1}$ values were obtained using the TPSS functional, however, for reasons that will become apparent below, we have chosen to use the PBEsol DF instead. The pressure derivative of the bulk modulus, $B_{1}$, were obtained by fitting 14 to 25 energy points evenly distributed around the equilibrium volume to the equation of state of Vinet {\it et al.}\cite{Vinet} We find that the ZPAE correction to the lattice constants is on average 0.019~\AA~with a maximum for LiF ($\Delta a_{0}=0.041$~\AA) and a minimum for InSb ($\Delta a_{0}=0.006$~\AA). The errors shown in Tables~\ref{tab:error} and \ref{tab:lattice-constants} are calculated using the corrected values for the lattice constants. In Table~\ref{tab:lattice-constants} we show both the corrected and uncorrected experimental values. In Table~\ref{tab:appendix}, we show the experimental Debye temperatures and the calculated $B_{1}$ values used in evaluating the corrections. 
\par
Overall, for any DF approximation the error in the lattice constants is less or equal to 1.4\%, except for the case of RPBE where the error is 2.6\%. It is clear from Table~\ref{tab:error} that the PBEsol is the best performing DF approximation regarding the lattice constants, irrespective of the error measure. The PBEsol is closely followed by the PBE0, HSE and AM05 DF approximations. All of these approximations have a MARE smaller than 0.6\%. The LDA and PBE approximations have a MARE of about 1\% and together with the RPBE and TPSS approximations are the functionals that either overestimate or underestimate the lattice constants compared to experiment for all systems. The LDA is the only approximation that yields too short lattice constants for all systems. We find, as was also discussed by Csonka  {\it et al.},\cite{Csonka2009} that if the ZPAE correction had not been applied this would have given a bias to systems that have a tendency of yielding too long lattice constants, i.e. making, e.g. the PBE and RPBE approximations more accurate than they are, while making the LDA less accurate. We also find that the LDA is more accurate than the PBE, which was also found by Csonka {\it et al.}\cite{Csonka2009} when analysing their smaller set of 10 non-metal solids. For their set of 14 metals the PBE performed better with a MARE of 1.2\% while the LDA had a MARE of 2.7\%.\cite{Csonka2009}
\par
In Fig.~\ref{fig:lattice-constants} we show the relative error for each individual system for the different DF approximations. It can be clearly seen that the LDA underestimates the lattice constants for all systems. Furthermore, the PBE, RPBE and TPSS approximations overestimates the lattice constants of all systems. The other approximations, only underestimate the lattice constant for a few systems, and then by a very small amount. We note that regarding the lattice constants some systems are more difficult to achieve good agreement between experiment and theory, e.g. SiC and MgO. LDA underbinds the most for the rock-salt structured crystals as well as for the two fluorite structured compounds.
\par

\begin{table*}[t]
\caption{\label{tab:elastic_cubic_c11} Elastic constant $c_{11}$ for cubic systems. Data shown in bold show the least deviation from experimental values. All values are given in GPa.}
\begin{ruledtabular}
\begin{tabular}{lccccccccc}
 & Expt.\cite{Landolt-Bornstein} & LDA & PBE & AM05 & PBEsol & RPBE & TPSS & PBE0 & HSE \\
\hline
C & 1079 & 1106 & 1056 & 1079 & 1075 & 1032 & 1052 & 1145 & 1143 \\
Si & 167 & 161 & 153 & 156 & 156 & 149 & 157 & 171 &  170 \\
Ge & 129 & 123 & 105 & 116 & 116 & 99 & 115 & 133 & 131 \\
BN & 820 & 826 & 781 & 796 & 795 & 1032 & 786 & 845 & 845 \\
BP & 315 & 358 & 339 & 346 & 346 & 330 & 340 & 368 & 367 \\
GaP & 140 & 142 & 124 & 133 & 134 & 117 & 126 & 143 & 142 \\
GaAs & 120 & 116 & 98 & 107 & 109 & 91 & 103 & 118 & 117 \\
InP & 100 & 100 & 87 & 94 & 94 & 81 & 87 & 103 & 101 \\
InAs & 83 & 85 & 71 & 77 & 79 & 65 & 72 & 87 & 86 \\
InSb & 67 & 67 & 54 & 60 & 61 & 50 & 56 & 68 & 67 \\
SiC & 390 & 404 & 384 & 390 & 390 & 375  & 388 & 418 & 417 \\
LiF & 136 & 166 & 117 & 119 & 134 & 94 & 121 & 132 & 131 \\
LiCl & 59 & 78 & 53 & 54 & 63 & 40 &  59 & 57 & 56 \\
MgO & 306 & 340 & 279 & 297 & 304 & 252 & 296 & 317 & 317 \\
NaF & 97 & 136 & 100 & 98 & 109 & 79 & 118 & 105 & 103 \\
CaF$_{2}$ & 164 & 191 & 159 & 164 & 171 & 144 & 160 & 171 & 170  \\
Mg$_{2}$Si & 121 & 125 & 117 & 120 & 120 & 113.5  & 118 & 128 & 127 \\
CoSb$_{3}$\cite{CoSb3-elastic} & 164 & 219 & 183 & 206 & 211 & 170 & 195 & 203 & 204 \\
\\
ME & -& 15.8 & -11.0 & -2.4 & \textbf{0.6} & -22.9 & -6.0 & 14.2 & 13.2 \\
MAE & -& 17.7 & 16.1 & 10.8 & \textbf{10.7} & 25.3 & 14.5 & 15.0 & 14.5 \\
RMSE & -& 24.3 & 18.5 & \textbf{15.5} & 15.9 & 29.3 & 17.3 & 24.1 & 23.7 \\
MRE & - & 9.4\%& -7.2\% & -2.6\% & \textbf{0.4\%} & -14.6\% & -3.4\% & 4.9\% & 4.1\%\\
MARE & -& 10.8\% & 9.7\% & 6.6\% & 6.5\% & 15.6\% & 8.7\% &  5.8\% & \textbf{5.5\%} \\
\end{tabular}
\end{ruledtabular}
\end{table*}

\begin{table*}[t]
\caption{\label{tab:elastic_cubic_c12} Elastic constant $c_{12}$ for cubic systems. Data shown in bold show the least deviation from experimental values. All values are given in GPa. Note that there is no experimental value for the $c_{12}$ elastic constant of CoSb$_{3}$. The value of $c_{12}$ can be evaluated to 45 GPa by using $c_{11}$ and  $B$ from Tables~\ref{tab:elastic_cubic_c11} and~\ref{tab:Bmod_cubic}, respectively, and by the use of Eq.~(\ref{eq:bulkmodcubic}).}
\begin{ruledtabular}
\begin{tabular}{lccccccccc}
 & Expt.\cite{Landolt-Bornstein} & LDA & PBE & AM05 & PBEsol & RPBE & TPSS & PBE0 & HSE \\
\hline
C & 124 & 149 & 125 & 141 & 141 & 115 & 122 & 138 &  138 \\
Si & 65 & 65 & 57 & 61 & 62 & 53 & 59 & 64 &  63 \\
Ge & 48 & 48 & 37 & 42 & 44 & 33 & 42 & 46 & 44 \\
BN & 190 & 191 & 168 & 181 & 182 & 115 & 173 & 186 & 185\\
BP & 100 & 85 & 73 & 80 & 81 & 68 & 73 & 79 & 79 \\
GaAs & 53 & 53 & 41 & 46 & 49 & 36 & 44 & 50 & 49  \\
GaP & 62 & 64 & 52 & 57 & 60 & 46 & 54 & 61 & 61  \\
InP & 56 & 57 & 46 & 50 & 53 & 41 &  47 & 55 & 54 \\
InAs & 45 & 48 & 37 & 42 & 44 & 33 & 40 & 46 & 45 \\
InSb & 37 & 37 & 28 & 31 & 34 & 24 & 29 & 35 & 34 \\
SiC & 142 & 143 & 128 & 138 & 138 & 121 & 132 & 141 & 141  \\
LiF & 47 & 52 & 49 & 45 & 47 & 45 & 46 & 50 & 50 \\
LiCl & 20 & 23 & 22 & 20 & 21 & 21 & 20 & 23 & 23 \\
MgO & 95 & 98 & 92 & 91 & 92 & 88 & 90 & 99 & 100 \\
NaF & 24 & 24 & 22 & 21 & 22 & 21 & 21 & 23 & 23 \\
CaF$_{2}$ & 44 & 61 & 41 & 43 & 48 & 32 & 41 & 46 & 45  \\
Mg$_{2}$Si & 22 & 27 & 23 & 25 & 25 & 21 & 24 & 25 & 25 \\
CoSb$_{3}$ & - & 47 & 37 & 42 & 46 & 33 & 38 & 33 & 27 \\
\\
ME & -& 3.0 & -7.8 & -3.5 & -1.8 & -12.9 & -6.9 & -0.5 & \textbf{-1.0} \\
MAE & -& 5.0 & 8.5 & 5.9 & 4.8 & 13.0 & 7.1 & \textbf{4.0} & 4.3 \\
RMSE & -& 8.6 & 11.2 & 7.9 & 7.0 & 15.8 & 9.6 & \textbf{6.5} & 6.8 \\
MRE & -& 6.3\% & -10.7\% & -5.9\% & -2.2\% & -18.9\% & -9.6\% & \textbf{0.3\%} & -0.9\%\\
MARE & -& 8.6\% & 13.0\% & 9.0\% & 7.2\% & 19.4\% & 10.5\% & \textbf{6.3\%} & 6.7\% \\
\end{tabular}
\end{ruledtabular}
\end{table*}
\begin{table*}[t]
\caption{\label{tab:elastic_cubic_c44} Elastic constant $c_{44}$ for cubic systems. Data shown in bold show the least deviation from experimental values. All values are given in GPa.}
\begin{ruledtabular}
\begin{tabular}{lccccccccc}
 & Expt.\cite{Landolt-Bornstein} & LDA & PBE & AM05 & PBEsol & RPBE & TPSS & PBE0 & HSE \\
\hline
C & 578 & 594 & 560 & 582 & 577 & 547 & 560 & 613 & 611 \\
Si & 80 & 76 & 74 & 74 & 74 & 74 & 77 & 82 &  82 \\
Ge & 67 & 64 & 56 & 61 & 61 & 54 & 61 & 69 & 68  \\
BN & 480 & 475 & 444 & 461 & 458 & 547 & 447 & 485 & 484\\
BP & 160 & 194 & 185 & 189 & 188 & 181 & 187 & 201 & 200 \\
GaP & 70 & 71 & 64 & 68 & 67 & 62 & 66 & 73 & 72 \\
GaAs & 60 & 58 & 51 & 55 & 55 & 48 & 53 & 60 & 60 \\
InP & 46 & 45.5 & 42 & 44 & 43 & 40 & 41 & 48 & 48 \\
InAs & 39 & 38 & 33 & 36 & 36 & 32 & 34 & 40 & 40 \\
InSb & 30 & 29 & 26 & 28 & 27 & 24 & 26 & 31 & 30 \\
SiC & 256 & 252 & 239 & 243 & 243 & 234 & 245 & 261 & 260   \\
LiF & 69 & 70 & 60 & 59 & 61 & 56 & 61 & 65 & 65 \\
LiCl & 27 & 28 & 25 & 24 & 25 & 23 & 26 & 26 & 26 \\
MgO & 157 & 156 & 145 & 148 & 148 & 140 & 146 & 158 & 158  \\
NaF & 28 & 29 & 27 & 28 & 28 & 26 & 30 & 29 & 30 \\
CaF$_{2}$ & 34 & 40 & 30 & 32 & 34 & 24 & 32 & 34 & 34   \\
Mg$_{2}$Si & 46 & 50 & 46 & 47 & 47 & 44 & 46 & 49 & 49 \\
CoSb$_{3}$\cite{CoSb3-elastic} & 59 & 57 & 50 & 54 & 53 & 48 & 51 & 60 & 60 \\
\\
ME & -& \textbf{2.3} & -7.2 & -2.8 & -3.4 & -10.9 & -5.5 & 5.6 & 5.1 \\
MAE & -& \textbf{4.9} & 9.9 & 6.6 & 6.6 & 13.2 & 8.7 & 6.0 & 5.6 \\
RMSE & -& \textbf{9.3} & 13.3 & 9.7 & 9.9 & 17.4 & 12.3 & 12.8 & 12.4 \\
MRE & -& \textbf{2.0\%} & -8.2\% & -4.4\% & -4.4\% & -12.2\% & -5.7\% & 3.6\% & 3.0\%\\
MARE & -& 4.9\% & 9.9\% & 6.6\% & 6.6\% & 13.6\% & 8.2\% & 4.4\% &  \textbf{3.8\%}\\ 
\end{tabular}
\end{ruledtabular}
\end{table*}

\subsection{Elastic constants}

As can be seen in Tables~\ref{tab:error}, \ref{tab:elastic_cubic_c11}, \ref{tab:elastic_cubic_c12} and \ref{tab:elastic_cubic_c44} the errors in the elastic constants   are larger, compared to the error in the lattice constants, with MARE of the order of 10\% or better depending on the functional. In general, the DF approximations underestimate the elastic constants. It is only the LDA that consistently overestimate the values for the elastic constants. The PBEsol, PBE0 and HSE approximations overestimate $c_{11}$ but underestimate both $c_{12}$ and $c_{44}$. We also find that the MAE and RMSE for $c_{11}$ are rather large, see Table~\ref{tab:elastic_cubic_c11}, so that the MAE and RMSE errors in $c_{11}$ is larger than the corresponding errors in both $c_{12}$ and $c_{44}$ for all DF approximations. In addition, both the MAE and RMSE result in a larger error in $c_{44}$ compared to $c_{12}$, except for the MAE for LDA which gives that the error in $c_{12}$ is about the same size as the error in $c_{44}$. However, the absolute error derived from the MAE or RMSE is not the best way to obtain a proper view of the performance of the DF approximations, since the size of the values for $c_{11}$, $c_{12}$ and $c_{44}$ for each system is very different; typically $c_{11}$ is much larger than both $c_{12}$ and $c_{44}$, and $c_{44}$ is larger than $c_{12}$. If we instead focus on the MARE, we find that the error in $c_{11}$ is smaller than the error in $c_{12}$ and the error in $c_{44}$ is smaller than the error in $c_{12}$. In general, for the MARE it can also be concluded that the error in $c_{11}$ is larger than the error in $c_{44}$, with the exception being the PBE, AM05 and PBEsol approximations. It is because of the large variations in the size of the errors in the different elastic constants we also evaluated the averaged errors as shown in Table~\ref{tab:error}. The averaged error estimates for the elastic constants shown in Table~\ref{tab:error} is intended to show the error that is expected for any single elastic constant when using any of the DF approximations. 
\par
If we are to determine which functional that performs the best, we note that it depends on which measure that is being used. In general, however, for $c_{11}$ we find that PBEsol is the best performing functional, followed by the two hybrid approximation and AM05. For $c_{12}$, the best performer is PBE0 followed by HSE, PBEsol and AM05. Interestingly, for $c_{44}$ LDA has the smallest ME, MAE, RMSE and MARE, while the HSE has the smallest MARE. For the averaged errors in Table~\ref{tab:error}, PBEsol has the smallest ME, MAE and MRE, AM05 has the smallest RMSE, while HSE has the smallest MRE and MARE. However, the differences between PBEsol, AM05 and the two hybrid functionals regarding the elastic constants are overall not very significant.
\par

\subsection{Bulk modulus}
It is common practise to evaluate the bulk modulus when comparing the performance of different density functional approximations.\cite{Mattsson,Csonka2009,Staroverov2004} The traditional method for determining the bulk modulus, $B$, is to calculate the total energy, $E$, as a function of volume, $V$, and fit the volume dependence of the energy to an equation of state. The bulk modulus at equilibrium can thereafter be obtained either as one of the fitting parameters or by evaluating
\begin{equation}
B = \frac{1}{V}\frac{\partial^2 E}{\partial V^2}.
\end{equation}
It is also possible to calculate the bulk modulus in terms of the elastic constants; for a cubic system it can be evaluated by\cite{Grimvall}
\begin{equation}\label{eq:bulkmodcubic}
B = \frac{c_{11}+2c_{12}}{3}.
\end{equation}
Here we have calculated the bulk modulus according to Eq.~(\ref{eq:bulkmodcubic}).
\par
We note that it is possible to perform a similar correction to the experimental bulk modulus as was done for the lattice constants.\cite{Alchagirov2001}~However, when considering the bulk modulus, as well as the elastic constants discussed in the previous section, it is necessary to point out that (i) the experimentally determined bulk modulus can have a measurement uncertainty in the order of 10\%, and (ii) the deviation between theory and experiment, as will be discussed, are much larger than for the lattice constants. Depending on the system and which DF approximations that is used the error varies from a couple of percent to about 20\%. The ZPAE correction is therefore of less importance for the bulk modulus and the elastic constants. We have therefore compared our theoretical data with experimental data without corrections. 
\par
The overall best performing functional is PBEsol, with the smallest ME, RMSE and MRE, while HSE has the smallest MAE and MARE. It is interesting to compare the error in the bulk modulus with the error in the elastic constants. In general, as shown in Table~\ref{tab:error}, the error in the elastic constants are larger than the error in the bulk modulus. The MAE and RMSE are both consistently larger for the elastic constants. The ME gives more or less similar error for both the elastic constants and the bulk modulus while there is no particular trend for the ME and MRE. The MARE is larger for the elastic constants, except for the LDA and RPBE approximations, where the error in the elastic constants is smaller than the corresponding error in the bulk modulus. However, if we look at the best performing functional, the error in the elastic constants are almost twice as large as the error in the bulk modulus, see Tables~\ref{tab:error},~\ref{tab:elastic_cubic_c11},~\ref{tab:elastic_cubic_c12},~\ref{tab:elastic_cubic_c44} and \ref{tab:Bmod_cubic}.
\par
The error in the bulk modulus can be written in terms of the errors in $c_{11}$ and $c_{12}$ by using Eq.~(\ref{eq:bulkmodcubic}):
\begin{equation}
\Delta B = \frac{1}{3}\Delta c_{11} + \frac{2}{3}\Delta c_{12}.
\end{equation}
From the above expression, it is clear that for a particular system the error in the elastic constants $c_{11}$ and $c_{12}$ will have different contributions to the error in the bulk modulus. An error in $c_{12}$ will have twice the contribution to the error in the bulk modulus compared to $c_{11}$. However, only one third of the error in $c_{11}$, and two thirds of the error in $c_{12}$, is transferred to the error in the bulk modulus. A large error in $c_{11}$ and a small error in $c_{12}$ will thereby results in an error in the bulk modulus that is smaller than the error in the elastic constants. The same is also true for the opposite case of a small error in $c_{11}$ and a large error in $c_{12}$. On the other hand, if $\Delta c_{11}\approx\Delta c_{12}$ this leads to $\Delta B\approx\Delta c_{11}$, i.e. the error in the bulk modulus is the same as the error in $c_{11}$. However, even for the cases were the errors in $c_{11}$ and $c_{12}$ are of similar size, the error in the bulk modulus is much smaller for the best performing functionals, such as PBEsol, AM05, HSE and PBE0. The reason for this lies in an overall cancellation of errors due to the overestimation of $c_{11}$ and underestimation of $c_{12}$ for the PBEsol, HSE and PBE0.
\par
\begin{table*}[t]
\caption{\label{tab:Bmod_cubic} Bulk modulus for cubic systems. Data shown in bold show the least deviation from experimental values. All values are given in GPa.}
\begin{ruledtabular}
\begin{tabular}{lccccccccc}
 & Expt.\cite{Landolt-Bornstein} & LDA & PBE & AM05 & PBEsol & RPBE & TPSS & PBE0 & HSE \\
\hline
C & 442 & 468 & 436 & 454 & 452 & 421 & 432 & 473 & 473\\
Si & 99 & 97 & 89 & 93 & 94 & 85 & 92 & 100 & 99\\
Ge & 75 & 73 & 60 & 67 & 68 & 55 & 66 & 75 & 73\\
BN & 400 & 403 & 373 & 386 & 387 & 420 & 377 & 405 & 405\\
BP & 172 & 176 & 162 & 169 & 169 & 155 & 162 & 175 & 175 \\
GaP & 88 & 90 & 76 & 82 & 85 & 70 & 78 & 89 & 88\\
GaAs & 75 & 74 & 60 & 66 & 69 & 54 & 64 & 73 & 71\\
InP & 71 & 71 & 59 & 65 & 67 & 54 & 61 & 71 & 70\\
InAs & 58 & 60 & 49 & 53 & 56 & 44 & 50 & 60 & 59\\
InSb & 47 & 47 & 37 & 41 & 43 & 33 & 38 & 46 & 45\\
SiC & 225 & 230 & 213 & 222 & 222 & 206 & 217 & 234 & 233\\
LiF & 77 & 90 & 71 & 70 & 76 & 62 & 71 & 78 & 77\\
LiCl & 33 & 41 & 32 & 31 & 35 & 27 & 33 & 34 & 34 \\
MgO & 165 & 179 & 154 & 160 & 163 & 143 & 159 & 172 & 172 \\
NaF & 49 & 61 & 48 & 47 & 51 & 40 & 54 & 51 & 50\\
CaF$_{2}$ & 84 & 104 & 81 & 84 & 89 & 69 & 81 & 88 & 87  \\
Mg$_{2}$Si & 55 & 60 & 55 & 56 & 57 & 21 & 55 & 60 & 59\\
CoSb$_{3}$ & 85 & 104 & 85 & 97 & 101 & 79 & 90 & 90 & 86\\
\\
ME & -& 7.2 & -8.9 & -3.1 & \textbf{-0.9} & -16.2 & -6.6 & 4.0 & 3.1 \\
MAE & -& 7.9 & 9.0 & 5.9 & 5.1 & 16.2 & 7.7 & 4.5 & \textbf{4.1} \\
RMSE & -& 10.9 & 11.1 & 7.0 & \textbf{6.5} & 18.1 & 9.2 & 8.2 & 8.0 \\
MRE & - & 8.0\% &-8.9\% & -4.1\% & \textbf{-0.7\%} & -16.6 & -5.9\% & 2.4\% & 1.2\%\\
MARE & -& 8.8\% & 8.9\% & 6.3\% & 5.3\% & 16.6\% & 7.8\% & 3.2\% & \textbf{2.7\%} \\
\end{tabular}
\end{ruledtabular}
\end{table*}

\section{Summary and Conclusions}\label{sec:conclusions}
We have performed DF calculations of the lattice constants, elastic constants and bulk modulus for a set of 18 semiconductors and insulators. We find, in agreement with previous studies, that the overall best performing functional is PBEsol, followed by the two hybrid approximations PBE0 and HSE, and AM05.  These functionals are distinct improvements over the LDA and PBE approximations. It should be kept in mind that the PBEsol and AM05 are from a calculation point of view much more efficient than PBE0 and HSE due to the very large computational cost of the hybrid functionals. If a reliable description of features of the electronic  structure, e.g. band gaps, is required it is necessary to use the hybrid functionals or to solve for the electronic structure by means of the GW approximation on top of a standard DF calculation. If structural and elastic properties are of interest, the PBEsol and AM05 functionals are the better choice for optimal efficiency, especially for large systems. Interestingly, the LDA performs better than the PBE for the lattice constants, bulk modulus and most of the elastic constants. It is only for $c_{11}$ were the error is smaller for the PBE approximation.
\par
The errors in the lattice constants are generally very small, less than 1.4\%. It is only RPBE which gives a larger error. The errors in the elastic constants and bulk modulus, on the other hand, are much larger compared to the error in the lattice constants, of about 10\% or smaller. Furthermore, we find that the error in the elastic constants are larger than the error in the bulk modulus. If the best performing functional is compared for both elastic constants and the bulk modulus, we find that the error in the elastic constants are about twice as large as the error in the bulk modulus. This is due to an overall cancellation of errors between the $c_{11}$ and $c_{12}$ elastic constants. 
\par
Finally, we note that the large deviations obtained using the RPBE should not discourage the use of this functional. It was designed for improving adsorption energies compared to the PBE and for such applications it is very successful.\cite{RPBE} 

\section{Acknowledgements}
We acknowledge support from the Leverhulme Trust via M. A. Moram's Research Leadership Award (RL-007-2012). M. A. Moram acknowledges further support from the Royal Society through a University Research Fellowship.

\appendix

\section{Additional data}
Here we provide the experimental Debye temperatures, $\Theta_{D}$, and calculated pressure derivatives, $B_{1}$, that was used in Eq.~(\ref{eq:correctionlattice}). The data are shown in Table~\ref{tab:appendix}.
\begin{table}[t]
\caption{\label{tab:appendix} Experimental Debye temperatures, $\Theta_{D}$, and calculated first pressure derivative of the Bulk modulus, $B_{1}$, for all systems in the present study.}
\begin{ruledtabular}
\begin{tabular}{lcc}
 & $\Theta_{D}$ (K)\cite{Landolt-Bornstein} & $B_{1}$\\
\hline
C & 2250 & 3.68 \\
Si  & 645 & 4.27 \\
Ge  & 373 & 4.70 \\
BN & 1700 & 3.67 \\
BP  & 985 & 3.78 \\
GaP  & 445 & 4.48\\
GaAs & 344 & 4.62\\ 
InP  & 425 &  4.67\\
InAs  & 280 & 4.78 \\
InSb  & 160 & 4.87\\
SiC  & 1200 & 3.90\\
LiF  & 732 & 4.33\\
LiCl  & 429 & 4.40 \\
MgO  & 945 & 4.15\\
NaF  & 430 & 4.72\\
CaF$_{2}$  & 510 & 4.56\\
Mg$_{2}$Si  & 417 & 4.04\\
CoSb$_{3}$  & 307 & 4.80\\
 \end{tabular}
\end{ruledtabular}
\end{table}

\bibliography{elastic}

\end{document}